\documentclass[english,aps,prl,showpacs,amsmath,amssymb,twocolumn]{revtex4}
\usepackage{epsfig}
\usepackage{bm}
\usepackage{bbold}
\usepackage{latexsym}

\begin{document}
	\title{Uncertainty principle on 3-dimensional manifolds \\of constant curvature}
	\author{Thomas Sch\"urmann}
	\email{<t.schurmann@icloud.com>}
	\affiliation{Germaniastra{\ss}e 8, 40223 D\"usseldorf, Germany}

\begin{abstract}
	We consider the Heisenberg uncertainty principle of position and momentum in 3-dimensional spaces of constant curvature $K$. The uncertainty of position is defined coordinate independent by the geodesic radius of spherical domains in which the particle is localized after a von Neumann-L\"uders projection. By applying mathematical standard results from spectral analysis on manifolds, we obtain the largest lower bound of the momentum deviation in terms of the geodesic radius and $K$. For hyperbolic spaces, we also obtain a global lower bound $\sigma_p\geq |K|^\frac{1}{2}\hbar$, which is non-zero and independent of the uncertainty in position. Finally, the lower bound for the Schwarzschild radius of a static black hole is derived and given by $r_s\geq 2\,l_P$, where $l_P$ is the Planck length.
\end{abstract}

\maketitle


\section{Introduction}
One of the open problems in contemporary physics is the unification of quantum mechanics and general relativity in the framework of quantum gravity. Quantum gravity phenomenology studies quantum gravity effects in low-energy systems. The basis of such phenomenological models is the generalized or the extended uncertainty principle (GUP/EUP) \cite{KMM95}\cite{BC05}\cite{P08}\cite{M10}. The main characteristics of such investigations typically consist of modified commutation relations between position and momentum, including a linear or quadratic dependence on the position or the momenta, as well as certain phenomenological parameters to highlight the terms originating from the linear and quadratic contribution related to the scale at which quantum-gravitational effects are expected to become relevant \cite{LP17}. However, it should be mentioned that both the GUP and the EUP are originally derived in the literature by using modified commutation relations introduced {\it ad hoc}. 

In \cite{CBL16}, a translation operator acting in a space with a diagonal metric has been introduced to develop a derivation of the EUP from first principles. It has been shown that for any (sufficiently smooth) metric expanded up to the second order, this formalism naturally leads to an EUP and to a minimum non-zero standard deviation of the momentum. This gives reason to expect the existence of even higher order corrections in the EUP if the metric had been considered without approximation.  

Rigorous mathematical derivations of uncertainty principles on Riemannian manifolds are hard to obtain. The problem already becomes apparent for quantum particles on the circle and on the sphere \cite{T04}. One of the difficulties for these systems is the position uncertainty measure for the particle (or the wave function spread measure). This is a consequence of the issue related to the choice of the operator for the azimuthal angle. This problem certainly holds for any closed manifold. For the 2-sphere the situation is even more complicated because of the absence of a self-adjoint momentum operator related to the azimuthal angle. This problem can be solved by the definition of a special coordinate system on the 2-sphere \cite{GP04}. 

In the following, we consider an alternative approach and focus on 3-dimensional spaces of constant sectional curvature $K$. Due to the theorem of Schur these are the spaces for which the Ricci tensor is proportional to the underlying metric $g$ (Einstein manifold).  
Our claim is to consider a physical situation which is sufficiently simple to be completely solvable but which still incorporates the main information contained in the interrelation between curvature and the uncertainty of position and momentum even for spaces with extreme values of curvature. 

Let us first consider the standard deviation of position $\sigma_{x_i}$, $i=1,2,3$, typically applied in the EUP. For spaces of positive constant curvature $K$, these measures are restricted by $\sigma_{x_i}\leq 1/\sqrt{K}$. Here, one might ask why we do not alternatively apply measures of uncertainty which also admit the total possible domain on the manifold. In fact, one might even claim that a measure of uncertainty in position should be able to cover the entire domain of the manifold, otherwise one could not be sure whether a certain implication (e.g. a cutoff) might just be a coordinate dependent artefact of an improperly defined uncertainty of position. Of course, this argument also applies for a standard deviation defined for the geodesics on a manifold.

To avoid these shortcomings, we consider an alternative measure of position uncertainty which will be defined by closed 3-dimensional balls $B_r$ of radius $r$ obtained by a standard von Neumann-L\"uders projection. The preliminary work of this approach has been given recently \cite{TS17}\cite{TS09}, where simultaneous measurements of position and momentum in flat (Euclidean) spaces of dimensions $n\geq 1$ have been discussed. 
More precisely, given a particle is strictly localized (prepared) in the domain $B_r\subset \mathbb{R}^n$, it has been shown that the standard deviation $\sigma_p$ of the canonical momentum is sharply bounded by the general inequality $\sigma_p \geq \lambda_1^{1/2}\hbar$, while $\lambda_1$ is the first Dirichlet eigenvalue of the Laplacian in $B_r$ and $\hbar=h/2\pi$ is the reduced Planck's constant of action. Of course, the eigenvalue $\lambda_1$ has to depend on $B_r$ and $r$ respectively. For the particular case of 3-dimensional Euclidean space, this statement can be explicitly expressed by  
\begin{eqnarray}\label{k0}
\sigma_p\,r \geq \pi\,\hbar.
\end{eqnarray} 
In the following section, we will go one step further and generalize this result to 3-dimensional spaces of sectional curvature $K\neq 0$. For the exceptional case of three dimensions there are closed form solutions of the corresponding eigenvalues and eigenfunctions. These will be derived in sections 3 and 4. Our main statement will be given in a proposition of section 5. Summary with outlook is given at the end. 

\section{The uncertainty principle on 3-manifolds of constant curvature}

A Riemannian manifold has three main notions of curvature: $(i)$ the Riemann curvature tensor (equivalent to the sectional curvature function, defined on tangent planes) which is a biquadratic form giving complete information at the curvature level; $(ii)$ the Ricci tensor $R_{ij}$, which is the trace of the Riemann tensor with respect to $g$; $(iii)$ finally, the Ricci scalar $S$, a function on $M$, which is the trace over the Ricci tensor with respect to $g$. 
	
When ${n = 2}$, the three curvatures are equivalent but the physical content of systems in two dimensions is somewhat restricted. When ${n = 3}$, the Ricci curvature still contains as much information as the Riemann curvature tensor and the physical relevance in this case is factual by the constant time section of the Robertson-Walker metric. The homogeneity at the curvature level leads us to the 3-dimensional manifolds of constant curvature. 
	
Let $(M,g)$ be a 3-dimensional simply connected Riemannian manifold $M$ with metric $g$ and constant sectional curvature $K$.
Under these assumptions, we consider a particle prepared in an interior ball $B_r\subset M$ by a quantum mechanical measurement process in terms of a standard von Neumann-L\"uders projection.  The uncertainty of the position in $M$ is then defined by the geodesic radius $r$ of the ball $B_r$. Accordingly, this definition is independent of the particular coordinate system. 

For a quantum mechanical system of this type the wave function of the particle is zero at the boundary $\partial B_r$. A Hilbert basis of $L^2(B_r)$, the space of square-integrable functions on $B_r$, is then defined by the Laplace-Beltrami operator on $B_r$ with Dirichlet boundary conditions \cite{CF78}:
\begin{eqnarray}\label{Lap}
\Delta \psi + \lambda\,\psi &=& 0\qquad\mbox{in }B_r,\\
\psi &=& 0\qquad\mbox{on } \partial B_r.
\end{eqnarray}
Let $\{\lambda_i\}$ be the set of eigenvalues and $\{u_i\}$ the orthonormal basis of eigenfunctions, $i=1,2,..$. It is known that there are an infinite number of eigenvalues with no accumulation point: $0<\lambda_1\leq\lambda_2\leq...$ and $\lambda_i\to\infty$ as $i\to\infty$. The scalar product in $L^2(B_r)$ will be denoted by angular brackets, that is to write $\langle \phi|\psi\rangle$ for two state vectors $\phi,\psi\in L^2(B_r)$. 

Now, we consider the standard deviation $\sigma_p$ of the momentum in $B_r$. For every wave function $\psi\in L^2(B_r)$ the eigenvalue problem (\ref{Lap}) is the same for its real part and its imaginary part. Both are co-linear and thus we only have to solve the real valued problem. According to (\ref{Lap}), the mean momentum of the particle in $B_r$ is zero and the square of  the momentum standard deviation is given by
\begin{eqnarray}\label{std1}
\sigma_p^2 = -\hbar^2 \langle\psi|\Delta\psi\rangle.
\end{eqnarray}
A sharp lower bound of $\sigma_p$ is now obtained by the first eigenvalue $\lambda_1$ of the Dirichlet eigenvalue problem which is in general dependent on the shape of the domain and we obtain the following inequality:
\begin{eqnarray}\label{UP}
\sigma_p \geq \lambda_1^\frac{1}{2}\hbar.
\end{eqnarray}
For a ball of arbitrary geodesic radius $r$ in a simply connected 3-manifold of constant curvature $K$, the first eigenvalue of the Laplace-Beltrami operator can explicitly be computed. In the following we give a brief description for positive and negative $K$.

\section{Spaces of constant positive curvature}
The universal covering of 3-manifolds of constant positive curvature is the 3-sphere. The 3-sphere $S^3$ of radius $R$ is the set of all points $x\in\mathbb{R}^4$ with $\langle x,x\rangle=R^2$ and $\langle\cdot,\cdot\rangle$ as the scalar product of the Euclidean 4-space. If we define the coordinates $(\chi,\theta,\varphi)$ by 
\begin{eqnarray}\label{spherical}
x&=&R \sin\chi\,\sin\theta\,\cos\varphi,\nonumber \\
y&=&R \sin\chi\,\sin\theta\,\sin\varphi,\nonumber \\
z&=&R \sin\chi\,\sin\theta\,\cos\varphi,\nonumber\\
w&=&R \cos\chi, \nonumber
\end{eqnarray}
for $0\leq\chi\leq\pi$, $0\leq\theta\leq\pi$, $0\leq\varphi\leq 2\pi$, then the (induced) metric on $S^3$ may be written as
\begin{eqnarray}\label{metricS3}
ds^2=R^2\left(d\chi^2+\sin^2\!\chi\,(d\theta^2+\sin^2\theta\, d\varphi^2)\right).
\end{eqnarray}
For our concern, it is obvious to consider the geodesic radius defined by $r=R\chi$, for which the metric takes the form 
\begin{eqnarray}\label{metricS3r}
ds^2=dr^2+\frac{\sin^2\!\sqrt{K} r}{K}\,(d\theta^2+\sin^2\theta\, d\varphi^2),
\end{eqnarray}
where $K=1/R^2$ is the curvature of the 3-sphere. The boundary value problem (\ref{Lap}) can now be expressed in the coordinates $(r,\theta,\varphi)$ corresponding to this metric. 
After separating the spherical harmonics (there is still the ordinary $SO(3)$ rotational invariance for any value of $K$), the relevant differential equation for the radial component $F(r)$ and for zero angular quantum numbers is given by   
\begin{eqnarray}\label{pde1}
& &F''+2\sqrt{K}\cot(\sqrt{K} r)F'+\lambda F =0, \quad K>0,\qquad\\
& &F(r_0)=0,\quad 0<r<r_0<\frac{\pi}{\sqrt{K}}.
\end{eqnarray}
The solution of the ordinary differential equation can be found in literature (cf.\,\cite{B83}\cite{P81}). The normalized eigenfunctions and the corresponding eigenvalues are given by 
\begin{eqnarray}\label{sol1}
& &F_n(r)=\sqrt{\frac{2 K}{r_0}}\,\,\frac{\sin \frac{n\pi}{r_0} r}{\sin\!\sqrt{K} r}\\
& &\lambda_n=\left(\frac{n\pi}{r_0}\right)^2-K,\qquad\qquad\qquad n=1,2,... \label{lS}
\end{eqnarray}
The normalization of the eigenfunctions is due to the corresponding volume measure 
\begin{eqnarray}\label{vol1}
dV_{S^3}=\frac{\sin^2\!\sqrt{K} r}{K}\,\sin\theta \,dr \,d\theta \,d\varphi.
\end{eqnarray}
Before we proceed, let us subsequently introduce the corresponding notation for the hyperbolic space.

\section{Spaces of constant negative curvature} 
For the case of negative curvature, the corresponding hyperbolic 3-space $H^3$ of radius $iR$ is the set of all points $x\in\mathbb{R}^{3,1}$ with 
$(x,x)=-R^2$ and $(\cdot,\cdot)=x^2+y^2+z^2-w^2$ as the Minkowski scalar product of the Lorentz 4-space.
It should be mentioned that $H^3$ cannot be isometrically immersed in the Euclidean 4-space - not even locally \cite{O54} -  with the consequence that the geodesic balls do not look spherically symmetric from the Euclidean point of view (in contrast to $S^3$ in $E^4$). However, in the Lorentz 4-space spherical symmetry is preserved for $H^3$ such that the first eigenvalue of the Laplace-Beltrami operator can be computed with respect to the circular domain around the "origin" of $H^3$ and is valid for every other circular domain of the same radius $r$ in $H^3$. Thus, we apply coordinates  $(\chi,\theta,\varphi)$ by 
\begin{eqnarray}\label{spherical}
x&=&R \sinh\chi\,\sin\theta\,\cos\varphi, \nonumber\\
y&=&R \sinh\chi\,\sin\theta\,\sin\varphi, \nonumber\\
z&=&R \sinh\chi\,\sin\theta\,\cos\varphi, \nonumber\\
w&=&R \cosh\chi, \nonumber
\end{eqnarray}
for $0\leq\chi<\infty$, $0\leq\theta\leq\pi$, $0\leq\varphi\leq 2\pi$ and obtain the corresponding metric on $H^3$ by 
\begin{eqnarray}\label{metricH3}
ds^2=R^2\left(d\chi^2+\sinh^2\!\chi\,(d\theta^2+\sin^2\theta\, d\varphi^2)\right).
\end{eqnarray}
Again we define the geodesic radius by $r=R\chi$, for which the hyperbolic metric takes the form 
\begin{eqnarray}\label{metricH3r}
ds^2=dr^2+\frac{\sinh^2\!\sqrt{|K|} r}{|K|}\,(d\theta^2+\sin^2\theta\, d\varphi^2),
\end{eqnarray}
where $K=-1/R^2$ is the corresponding curvature of $H^3$. Similar to the approach of the previous section, the relevant differential equation for the radial component $G(r)$ and for zero angular quantum numbers is 
\begin{eqnarray}\label{pde2}
&&G''+2\sqrt{|K|}\coth(\sqrt{|K|} r)G'+\lambda G =0, \quad K<0,\qquad\\
&&G(r_0)=0,\quad 0<r<r_0<\infty.
\end{eqnarray}
The solution of the ordinary differential equation is also known in literature \cite{S09}\cite{A16}, and can be written as 
\begin{eqnarray}\label{sol2}
& &G_n(r)=\sqrt{\frac{2 |K|}{r_0}}\,\,\frac{\sin \frac{n\pi}{r_0} r}{\sinh\!\sqrt{|K|} r}\\
& &\lambda_n=\left(\frac{n\pi}{r_0}\right)^2-K,\qquad\qquad\qquad n=1,2,... \label{lH}
\end{eqnarray}
The different meaning of the geodesic radius $r$ for the spherical and hyperbolic case will be clear from the context such that no additional notation will be necessary.

\section{Uncertainty principle on 3-dimensional Einstein manifolds}
\begin{figure}[t]
	\begin{center}
		\psfig{file=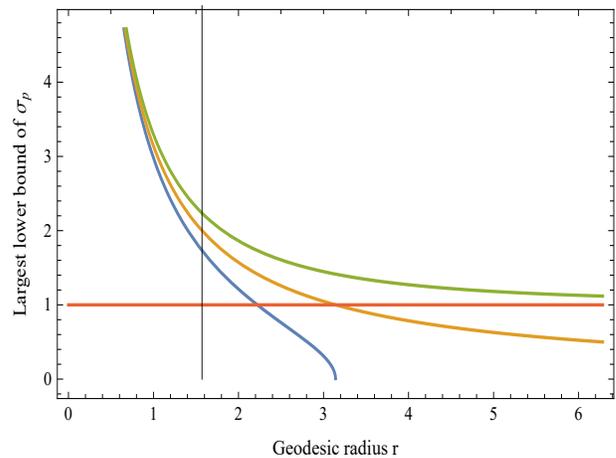,width=8.0cm, height=6.0cm}
		\parbox{8.0cm}{
			\caption{Largest lower bound of $\sigma_p$ (in units of $\hbar$) versus geodesic radius for $K=1$, $r\in(0,\pi]$  (blue), $K=0$ and $K=-1$, $r\in(0,\infty)$  (orange and green). Also shown is the minimum standard deviation for the hyperbolic case (red). The vertical line at $\pi/2$ (black) denotes the ”equator” of the 3-sphere for $K = 1$.   
				\label{fig1}}}
	\end{center}
\end{figure}

After substitution of the results (\ref{lS}) and (\ref{lH}) for ${n=1}$ into expression (\ref{UP}), we can summarize the following main result:\\
\\
\noindent$\bf{Proposition.}$ For a particle whose position is strictly localized in a closed circular subset $B_r$ of a 3-dimensional simply connected space of constant curvature $K$, we obtain the following representation of the uncertainty principle
\begin{eqnarray}\label{K}
\sigma_p\, r\geq\pi\hbar \left[1-\frac{K}{\pi^2}\, r^2\right]^\frac{1}{2},
\end{eqnarray}
with the geodesic radius $r\in[0,\pi/K^{1/2}]$ for $K\geq 0$, or otherwise $r\in[0,\infty)$, if $K< 0$. The equal sign in (\ref{K}) is obtained for the first eigenfunctions in (\ref{sol1}) and (\ref{sol2}) respectively. \\
\\
Although this closed form expression is valid for all values of $K$, we want to note that the corresponding physical context for ${K>0}$ and ${K<0}$ is very different. For the simple case ${K=0}$ we obtain expression (\ref{k0}), which has already been discussed in \cite{TS17}. The other two cases are as follows:\\   
\\
{$\underline{\bf K>0:}$} The corresponding space is $S^3$ with standard metric induced by the Euclidean 4-space. In this space, the volume of the ball with position uncertainty corresponding to the geodesic radius $r$ is 
\begin{eqnarray}\label{vol1}
V_+(r)=2 \pi r R^2\left(1-\frac{\sin(2r/R)}{2r/R}\right),
\end{eqnarray}
while the total domain of measure $2\pi^2 R^3$ is reached for the maximum possible value $r\to\pi R$ (the maximal length in $S^3$). In this case, the right-hand side of (\ref{K}) approaches zero and the momentum dispersion can be equal to zero while the position uncertainty is still finite (see FIG.\,\ref{fig1}). Actually, this special behaviour confirms the situation for the uncertainty principle on closed manifolds mentioned in \cite{GP04}. In the limit $R\to \infty$, the volume $V_+$ approaches the ordinary volume of a ball in Euclidean space corresponding to the situation discussed in \cite{TS17}. \\
\\
{$\underline{\bf K<0:}$} The corresponding space is $H^3$ with the standard metric induced by the Lorentzian 4-space. The ball with position uncertainty corresponding to radius $r$ has the volume 
\begin{eqnarray}\label{vol2}
V_-(r)=2 \pi r R^2\left(-1+\frac{\sinh(2 r/R)}{2 r/R}\right).
\end{eqnarray}
For $R\to \infty$, this subset also approaches the standard volume of a ball in Euclidean space. A remarkable fact is given when the position uncertainty $r$ tends to infinity while $R$ is kept fixed. Then, we obtain the following lower bound (FIG.\,\ref{fig1})
\begin{eqnarray}\label{K1}
\sigma_p\geq |K|^\frac{1}{2}\hbar.
\end{eqnarray}
In contrast to the situation for the EUP, here the position uncertainty is not simultaneously bounded from above. 
On the other hand, for $R\to 0$, the standard deviation of the momentum tends to infinity, which might be hard to accept from the physical point of view. 

In general, for small values ${r/R\ll 1}$, the first order terms in the Taylor series of (\ref{K}) are given by 
\begin{eqnarray}\label{K2}
\sigma_p\geq \pi\hbar\left(\frac{1}{r}-\frac{K}{2\pi^2}\,r\right)+{\cal O}(K^2r^3)
\end{eqnarray}
and the term in the brackets clearly remembers the formal structure of the EUP, where the pre-factor of $r$ might be associated with the phenomenological parameter $\alpha$ in\cite{KMM95} \cite{LP17} or $\beta$ in \cite{P08}. At this point some caution should be exercised, because the qualitative behaviour of the exact result (\ref{K}) and the  terms in (\ref{K2}) is quite different. When ${K<0}$,  then expression (\ref{K2}) reaches its absolute minimum already for a finite value $r=\pi\sqrt{2/|K|}$ and not only in the limit $r\to\infty$. For $K>0$, the term in brackets is zero when $r=\pi R \sqrt{2}$ which is beyond the admissible domain of $r$. 
Therefore, it is important to be cautious whether the inequality of the EUP is considered to be a first order Taylor series approximation (c.f.\,\cite{BU08}) or alternatively the square root of the type given in (\ref{K}). \\

By applying (\ref{K}), we are in the position to give a rigorous derivation of the Planck length $l_P=\sqrt{\hbar G/c^3}$. Consider the Schwarzschild radius of a particle with rest mass $m$. In our notation (\ref{std1}), the relativistic energy-momentum relation of that particle satisfies the inequality ${E\geq c\,\sigma_p}$, while the equal sign is given for $m=0$. 
The Schwarzschild radius of the particle is $r_s=2 G M/c^2$, while the mass $M$ is given by the energy of the particle according to ${M=E/c^2}$. It follows that the Schwarzschild radius is bounded from below according to the inequality 
\begin{eqnarray}\label{rs1}
r_s\geq \frac{2 G}{c^3}\,\sigma_p.
\end{eqnarray}
Any object whose extension is smaller than the sphere corresponding to its Schwarzschild radius is called a black hole. When the particle is completely localized inside the spherical domain of radius $r_s$, we can regard the lower bound given in (\ref{K}), while the corresponding geodesic radius is given by 
\begin{eqnarray}\label{rs2}
r=\int_{0}^{r_s}\!d\tilde{r}\,\left|1-\frac{r_s}{\tilde{r}}\right|^{-\frac{1}{2}}=    \pi r_s/2. 
\end{eqnarray}
Because of the asymptotic flatness of the Schwarzschild solution and since the Ricci scalar is zero, we choose $K=0$ such that the lower bound of the standard deviation is $\sigma_p\geq 2\hbar/r_s$. It follows that the Schwarzschild radius of the black hole has to satisfy the condition 
\begin{eqnarray}\label{rs}
r_s\geq 2 \,l_P.
\end{eqnarray} 
This lower bound confirms the (heuristic) derivations typically proposed in literature. Of course, there are objects with Schwarzschild radii much smaller than those classified in (\ref{rs}), but these objects are not completely localized inside the domain corresponding to $r_s$ and therefore not black holes. The claim of our brief derivation is that any Schwarzschild black hole must satisfy the condition (\ref{rs}). \\

\section{Summary and outlook}

Based on standard results of mathematical spectral analysis on manifolds, we introduced an uncertainty relation on 3-dimensional manifolds of constant curvature. This result can be considered as an interrelation between the coordinate independent defined uncertainties of position and momentum on the one hand, and the curvature $K$ as a coordinate invariant property on the other hand. Because no approximation has been applied in the derivation of the result (\ref{K}), it can be applied without any restriction concerning the domain of definition. This advantage is not self-evident, but is caused by the fact that a closed form expression for the first Dirichlet eigenvalue on curved manifolds is only given in three dimensions. For higher dimensions there are also estimates in literature, but these are not sharp. 

If one thinks that constancy of the sectional curvature is too strong, we might consider manifolds of constant scalar curvature. Here, it is known that on any compact manifold there exist Riemannian metrics of constant scalar curvature and they form an infinite dimensional family even in three dimensions. Therefore, one alternatively consider manifolds whose Ricci tensor satisfies the condition ${Ric\geq (n-1)\,k\,g}$, for some non-negative constant $k$. Under this assumption, there is a result given by Reilly \cite{R77}, that $\lambda_1\geq n k$, if the corresponding domain has a weakly convex boundary. At first glance this bound seems to be independent of the position uncertainty. However, a dependence on the position is implicitly contained in the weakly convex boundary assumption. For instance, in the trivial case of $S^3$, this bound is valid only for domains restricted to the "upper hemisphere", that is when $r\in[0,\pi R/2)$. 
Another estimate, valid under the same assumptions but which explicitly depends on $r$ as well, has been derived recently in\cite{L06}. Even though this estimate is mathematically hard to prove, it is still not sharp and thus also of restricted applicability. However, the search for improved estimates in this direction is part of present activities in mathematical research.\\
\\

\end{document}